\documentclass[twocolumn,prb]{revtex4}
\usepackage{amsfonts}
\usepackage[T1]{fontenc}
\usepackage{amsmath,amsbsy,amssymb,graphicx}
\usepackage{times}

\begin{document}

\title{Photoinduced topological phase transition\\
from a crossing-line nodal semimetal to a multiple-Weyl semimetal}
\author{Motohiko Ezawa}
\affiliation{Department of Applied Physics, University of Tokyo, Hongo 7-3-1, 113-8656,
Japan}

\begin{abstract}
We propose a simple scheme to construct a model whose Fermi surface is
comprised of crossing-line nodes. The Hamiltonian consists of a normal
hopping term and an additional term which is odd under the mirror
reflection. The line nodes appear along the mirror-invariant planes, where
each line node carries the quantized Berry magnetic flux. We explicitly
construct a model with the $N$-fold rotational symmetry, where the $2N$ line
nodes merge at the north and south poles. Photoirradiation induces a
topological phase transition. When we apply photoirradiation along the $k_z$
axis, there emerge point nodes carrying the monopole charge $\pm N$ at these
poles, while all the line nodes disappear. The resultant system describes
anisotropic multiple-Weyl fermions.
\end{abstract}

\maketitle

\textit{Introduction:} Weyl semimetal is one of the hottest topics in
condensed matter physics\cite{Hosur,Jia}. It is protected by a monopole
charge in the momentum space\cite{Murakami}. Multiple Weyl semimetal is a
generalization of a Weyl semimetal which has a monopole charge larger than
the unit charge\cite{CFang,BJYang,Xli,Huang2}.There exist another class of
novel semimetals. They are line nodal semimetals whose Fermi surfaces form
one-dimensional lines\cite%
{Burkov,CFangA,Xie,Yamakage16,Hyper,Hirayama,Sy,Carter,Phillips,ChenLu,Chiu,Mullen,Weng,Bian,Rhim,ChenXie,Fang,BianChang}%
. A line node is protected by a quantized Berry magnetic flux. Recently,
line nodal semimetals are generalized into two species. One is a loop node
forming a nontrivial link such as the Hopf link\cite{WChen,ZYan,PYChang,Hopf}%
. The other is a crossing-line node, where several line nodes cross at a
point\cite{Zeng,Chain,Kim,Yamakage,CaTe,Yu}.

Photoirradiation is a powerful tool to modify the band structure\cite%
{Oka09L,Kitagawa01B,Lindner,Dora,EzawaPhoto,Gold}. According to the Floquet
theory an additional term emerges due to the second-order process of
photoirradiation. A typical example is the generation of a Weyl point from a
Dirac semimetal\cite{PWang,PChan, Ebihara,PChan2}. It is shown that a Weyl
node can also be generated by applying photoirradiation to a loop nodal
semimetal\cite{PYan,Multi}.

In this paper, we first propose a simple scheme to construct models for
crossing-line nodal semimetals. We then investigate how a crossing-line
nodal semimetal is modified by way of photoirradiation. The model
Hamiltonian consists of a normal hopping term and a mirror-odd interaction
term. A line node emerges on the mirror-invariant plane. Each line node is
topologically protected by the quantized Berry magnetic flux. Explicitly, we
construct an $N$-fold rotational symmetric model, where the crossing of $2N$%
-fold line nodes occurs at the north pole and also at the south pole. There
are no magnetic monopoles at these poles. Next, we derive a photoinduced
term based on the Floquet theory. It induces a topological phase
transition.\ Indeed, by applying photoirradiation along the $z$ direction,
the Fermi surface is found to disappear by the emergence of gap except for
two nodal points carrying the $N$ ($-N$) units of the monopole charge at the
north (south) pole. The resultant system is the anisotropic $N$-fold
multiple-Weyl semimetal. On the other hand, by applying photoirradiation
perpendicular to one of the loop node, the resultant Fermi surface turns out
to be comprised of this loop node and point nodes. Otherwise, only the point
nodes appear.

\textit{Model:} A prototype of line nodal semimetals is given by the model 
\begin{equation}
H(\mathbf{k})=f_{x}(\mathbf{k})\sigma _{x}+f_{z}(\mathbf{k})\sigma _{z},
\label{BasicHamil}
\end{equation}%
whose energy spectrum reads%
\begin{equation}
E(\mathbf{k})=\pm \sqrt{f_{x}^{2}(\mathbf{k})+f_{z}^{2}(\mathbf{k})}.
\end{equation}%
The Fermi surface is given by the two conditions $f_{x}(\mathbf{k})=0$ and $%
f_{z}(\mathbf{k})=0$, each of which produces a two-dimensional surface. The
intersection of the two surfaces consists of lines and/or points in general.
Namely we obtain line nodes and/or point nodes in general. For simplicity we
consider the following case: (i) The condition $f_{x}(\mathbf{k})=0$
generates an ellipsoid, which is rotational symmetric around the $k_z$ axis
and centered at the origin ($\mathbf{k}=0$). (ii) The condition $f_{x}(%
\mathbf{k})=0$ generates planes sharing the $k_z$ axis with each other.
Furthermore we require that $f_{z}(\mathbf{k})$ is odd under the mirror
operation $M_{\alpha }$ with respect to each plane, where the index $\alpha $
denotes the direction normal to the plane. For example, if $f_{z}(\mathbf{k}%
) $ is odd for the mirror reflection with respect to the $k_{y}k_{z} $
plane, $M_{x}f_{z}(k_{x},k_{y},k_{z})M_{x}^{-1}=-f_{z}(-k_{x},k_{y},k_{z})$,
we have a zero-energy solution at $k_{x}=0$ since $%
M_{x}f_{z}(0,k_{y},k_{z})M_{x}^{-1}=-f_{z}(0,k_{y},k_{z})$.

\begin{figure*}[t]
\centerline{\includegraphics[width=0.95\textwidth]{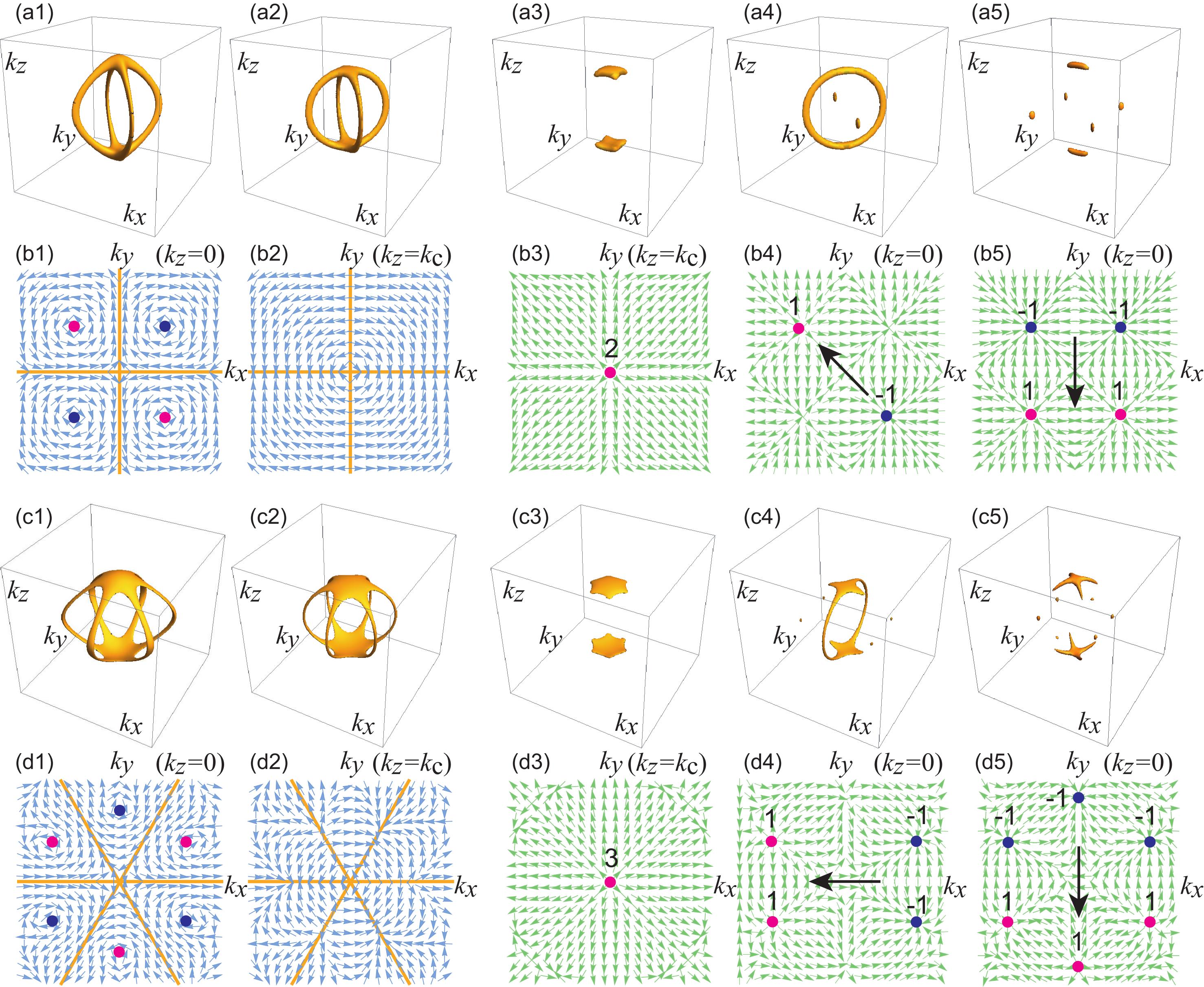}}
\caption{Bird's eye's view of the almost zero-energy surfaces of the
Hamiltonian with $N=2$ for (a1) the lattice model, (a2) the continuum model,
(a3) the continuum model with photoirradiation along the $k_{z}$ axis,
(a4)--(d5) the continuum model with photoirradiation perpendicular to the $%
k_{z}$ axis, where the direction is indicated by arrows as in (b4)--(b5).
(b1) Normalized Berry connection $(A_{x},A_{y})/\protect\sqrt{%
A_{x}^{2}+A_{y}^{2}}$ for the lattice model with $N=2$ on the $k_{z}=0$
plane. Red (blue) dots represent vortices and antivortices. The total
vorticity is zero. (b2) Normalized Berry connection for $N=2$ on the $k_{z}=k_{c}$
plane, where the north pole is present at the center of the plane. There are
no vortices and no monopoles there. (b3) Normalized Berry curvature $%
(B_{x},B_{y})/\protect\sqrt{B_{x}^{2}+B_{y}^{2}}$ corresponding to (a3) on
the $k_{z}=k_{c}$ plane. (b4)--(b5) Normalized Berry curvature corresponding
to (a4)--(a5) on the $k_{z}=0$ plane. Red (blue) dots represent monopoles
and antimonopoles carrying the monopole charges indicated by the attached
numbers. No monopoles appear at the north and south poles. (c1)--(c5) The
corresponding bird's eye's view for $N=3$. (d1)--(d2) The corresponding
normalized Berry connection for $N=3$. (d3)--(d5) The corresponding
normalized Berry curvature for $N=3$, where the direction of the
photoirradiation is indicated by arrows. }
\label{FigBerry}
\end{figure*}

A key observation is that, when there are $N$ mirror-odd planes, there
emerges the crossing of $2N$ line nodes. For example, by assuming the $N$%
-fold rotation symmetry, the lattice Hamiltonian is given by (\ref%
{BasicHamil}) together with%
\begin{align}
f_{x}(\mathbf{k})& =t\sum_{j=1}^{N}\cos \left( \mathbf{d}_{j}^{c}\cdot 
\mathbf{k}\right) +t_{z}\cos k_{z}-m^{\prime },  \notag \\
f_{z}(\mathbf{k})& =\lambda ^{\prime }g\left( k_{z}\right)
\prod\limits_{j=1}^{N}\sin \left( \mathbf{d}_{j}^{s}\cdot \mathbf{k}\right)
,  \label{EqA}
\end{align}%
where $\mathbf{d}_{j}^{c}=\left( \cos \left[ j\pi /N\right] ,\sin \left[
j\pi /N\right] ,0\right) $ and $\mathbf{d}_{j}^{s}=\left( \sin \left[ \left(
2j+1\right) \pi /\left( 2N\right) \right] ,\cos \left[ \left( 2j+1\right)
\pi /\left( 2N\right) \right] ,0\right) $. We have included a function $%
g\left( k_{z}\right) $ to allow the freedom of introducing additional
crossing-line modes perpendicular to the $k_{z}$ axis such as in (\ref{EqB})
for the cubic symmetric model. The summation $\sum_{j=1}$\ runs over the
nearest neighbor sites. The Fermi surface is given by the cross section of
the $N$ planes and the ellipsoid. They are $2N$ line nodes which cross at
the north and south poles. We show Fermi surfaces for $N=2$ and $N=3$ in
Figs.\ref{FigBerry}(a1) and (b1), respectively. (Actually, we present almost
zero-energy surfaces $E=\delta $ with $0<\delta \ll t$.) We note that the
lattice structure is possible in the real space only for $N=2$\ and $3$. The
lattice with $N=2$\ forms a layered square lattice, while the lattice with $%
N=3$\ forms a layered triangular lattice.\textbf{\ }Nevertheless, we analyze
the general $N$\ case to make the mathematical structure clearer.

The corresponding continuum theory is given by

\begin{equation}
H=[a\left( k_{x}^{2}+k_{y}^{2}\right) +ck_{z}^{2}-m]\sigma _{x}+\lambda
g\left( k_{z}\right) \text{Re}(k_{+}^{N})\sigma _{z}  \label{HamilConti}
\end{equation}%
with $a=-Nt/4$, $c=-t_{z}/2$, $m=m^{\prime }+Nt+t_{z}$ and $\lambda =\lambda
^{\prime }/2^{N-1}$.\textbf{\ }For example, for the case of $N=2$, there are
two mirror planes $M_{x+y}$ and $M_{x-y}$. A simplest representation is $%
f_{z}=k_{x}^{2}-k_{y}^{2}$, whose zero-energy solution is given by the two
planes $k_{x}=k_{y}$ and $k_{y}=-k_{y}$. For the case of $N=3$, there are
three mirror planes determined by $f_{z}=k_{x}^{3}-3k_{x}k_{y}^{2}$. We show
the Fermi surfaces in Figs.\ref{FigBerry}(a2) and (b2). The Fermi surfaces
obtained in the continuum theory are found to be almost the same as those
obtained in the lattice model. Hence, we use the continuum theory in the
following.

With the use of the eigenfunction $\left\vert \psi \right\rangle $ of the
Hamiltonian (\ref{HamilConti}) we may calculate the Berry connection as%
\begin{equation}
A_{i}\left( \mathbf{k}\right) =-i\left\langle \psi \right\vert \partial
_{i}\left\vert \psi \right\rangle =\frac{f_{x}\partial
_{i}f_{z}-f_{z}\partial _{i}f_{x}}{2\left( f_{x}^{2}+f_{z}^{2}\right) }=%
\frac{1}{2}\partial _{i}\Theta 
\end{equation}%
with $\partial _{i}=\partial /\partial k_{i}$, where $f_{x}=f\cos \Theta $, $%
f_{z}=f\sin \Theta $ and $f=\sqrt{f_{x}^{2}+f_{z}^{2}}$. We show the stream
plot of the Berry connection for $N=2$ and $3$, where vortex and antivortex
structures are observed around the line nodes in the constant $k_{z}$ plane,
as in Figs.\ref{FigBerry}(b1) and (d1). A pair of vortex and antivortex
annihilates at the north and south poles as in Figs.\ref{FigBerry}(b2) and
(d2). Each line node is topologically protected since the Berry phase along
the line nodes is quantized to be $\pm \pi $,%
\begin{equation}
\oint A_{j}dk_{j}=\int \nabla \times \mathbf{A}\,dS=\pm \pi .
\end{equation}%
The Berry curvature $\mathbf{B}=\nabla \times \mathbf{A}$ is strictly
localized along the line node. Indeed, we can explicitly check this by the
direct calculation,%
\begin{equation}
B_{i}\left( \mathbf{k}\right) =\varepsilon _{ijk}\partial _{j}A_{k}=\pm \pi
\sum \delta \left( f_{x}\right) \delta \left( f_{y}\right) .
\end{equation}%
Namely, the Berry magnetic flux is present along each line node, while the
Berry curvature is strictly zero away from the line nodes. Consequently a
line node is topologically protected.

\textit{Photoirradiation parallel to the }$k_{z}$ axis\textit{:} We proceed
to investigate a topological phase transition due to the $\sigma _{y}$ term
induced by photoirradiation. The following formulas hold for any function $%
g(k_{z})$ in (\ref{EqA}). We summarize the Floquet theory on
photoirradiation. First, we irradiate a beam of circularly polarized light
along the $z$ direction. We take the electromagnetic potential as $\mathbf{A}%
_{\text{EM}}(t)=(A\cos \omega t,A\sin \omega t,0)$, where $\omega $ is the
frequency of light with $\omega >0$ for the right circulation and $\omega <0$
for the left circulation. The effective Hamiltonian due to the second order
process of photoirradiation is given by\cite%
{Oka09L,Kitagawa01B,Lindner,Dora,EzawaPhoto,Gold}%
\begin{equation}
\Delta H_{\text{eff}}\left( \mathbf{k},\mathbf{A}_{\text{EM}}\right) =\frac{1%
}{\hbar \omega }\left[ H_{-1}\left( \mathbf{k},\mathbf{A}_{\text{EM}}\right)
,H_{+1}\left( \mathbf{k},\mathbf{A}_{\text{EM}}\right) \right] .
\end{equation}%
It is explicitly evaluated to be%
\begin{equation}
\Delta H_{\text{eff}}\left( \mathbf{k},\mathbf{A}_{\text{EM}}\right)
=f_{y}\sigma _{y},
\end{equation}%
where%
\begin{equation}
f_{y}=-2na\lambda \alpha g\left( k_{z}\right) \text{Im}(k_{+}^{N})
\end{equation}%
with $\alpha =\left( eA\right) ^{2}/\left( \hbar \omega \right) $. The
second-order perturbed process produces the term $f_{y}(\mathbf{k})\sigma
_{y}$ due to the commutation relation $[\sigma _{z},\sigma _{x}]=i\sigma
_{y} $. By including this term into the Hamiltonian we find%
\begin{equation}
H(\mathbf{k})=f_{x}(\mathbf{k})\sigma _{x}+f_{y}(\mathbf{k})\sigma
_{y}+f_{z}(\mathbf{k})\sigma _{z},
\end{equation}%
while the energy is modified as%
\begin{equation}
E(\mathbf{k})=\pm \sqrt{f_{x}^{2}(\mathbf{k})+f_{y}^{2}(\mathbf{k}%
)+f_{z}^{2}(\mathbf{k})}.
\end{equation}%
Now the condition $f_{y}(\mathbf{k})=0$ should be imposed as the zero-energy
condition additionally to $f_{x}(\mathbf{k})=f_{z}(\mathbf{k})=0$. In
general, there is no intersection between three surfaces, and the system
becomes an insulator. However, there are several cases where crossing-line
nodes are reduced to points nodes, as shown in Figs.\ref{FigBerry}(a3) and
(c3).

The Fermi surface consists of only two point nodes at the north and south
poles, $(k_{x},k_{y},k_{z})=(0,0,k_{\text{c}})$ with $k_{\text{c}}=\pm \sqrt{%
m/c}$. In the vicinity of these points, we have%
\begin{equation}
f_{z}\approx \pm 2\sqrt{mc}\left( k_{z}\mp \sqrt{m/c}\right) .
\end{equation}%
The Hamiltonian with photoirradiation is given by%
\begin{align}
H& =\pm 2\sqrt{mc}\left( k_{z}\mp \sqrt{m/c}\right) \sigma _{x}+\lambda
g\left( k_{z}\right) \text{Re}(k_{+}^{N})\sigma _{z}  \notag \\
& -2na\lambda \alpha g\left( k_{z}\right) \text{Im}(k_{+}^{N})\sigma _{y}.
\label{PhotoH}
\end{align}%
In particular, when $2na\left( eA\right) ^{2}=\hbar \omega $ and $g\left(
k_{z}\right) =1$, the Hamiltonian is reduced to that of the multiple-Weyl
fermion,%
\begin{equation}
H=\lambda \left( k_{+}^{n}\sigma _{+}+k_{-}^{n}\sigma _{-}\right) \pm 2\sqrt{%
mc}\left( k_{z}\mp \sqrt{m/c}\right) \sigma _{z},
\end{equation}%
and otherwise it is that of the anisotropic multiple-Weyl fermion. The Berry
curvature is calculated as%
\begin{align}
F_{i}\left( \mathbf{k}\right) & =\frac{\varepsilon _{ijk}}{2}\sin \Theta
\left( \partial _{j}\Theta \partial _{k}\Phi -\partial _{k}\Theta \partial
_{j}\Phi \right)   \notag \\
& =\varepsilon _{ijk}\left( \partial _{j}\mathbf{f}\times \partial _{k}%
\mathbf{f}\right) \cdot \mathbf{f}
\end{align}%
where $\mathbf{f}=\left( f_{x},f_{y},f_{z}\right) $ with $f_{x}=f\cos \Phi \sin
\Theta $, $f_{y}=f\sin \Phi \sin \Theta $, $f_{z}=f\cos \Theta $. It
describes monopoles with the charges $\pm N$ at the north and south poles.
We illustrate the Berry curvature around the north pole for $N=2$ and $3$ in
Figs.\ref{FigBerry}(b3) and (d3) for the case of $g\left( k_{z}\right) =1$,
where the presence of the monopoles is observed as a source or a sink of the
Berry magnetic flux. We conclude that, by applying photoirradiation along
the $z$ direction, the Fermi surface changes from the nodal crossing lines
to the two nodal points carrying the $N$ ($-N$) units of the monopole charge
at the north (south) pole.

\textit{Photoirradiation perpendicular to the }$k_{z}$ axis\textit{:} We
next apply photoirradiation perpendicular to the $k_{z}$ axis. We explicitly
study the tetragonal symmetric model ($N=2$) and the trigonal symmetric
model ($N=3$). Here we set $g(k_{z})=1$ in (\ref{EqA}).

The lattice Hamiltonian of the tetragonal symmetric model with $N=2$ is
given by 
\begin{align}
f_{x}& =t\left( \cos k_{x}+\cos k_{y}\right) +t_{z}\cos k_{z}-m^{\prime }, 
\notag \\
f_{z}& =-\lambda \left( \cos k_{x}-\cos k_{y}\right) .
\end{align}%
We inject photoirradiation along the $\phi $ direction with $\mathbf{A}_{%
\text{EM}}(t)=(-A\sin \phi \cos \omega t,A\cos \phi \cos \omega t,A\sin
\omega t)$. The effective Hamiltonian induced by photoirradiation along the $%
\phi $ direction is given in the continuum theory by 
\begin{equation}
f_{y}=-\alpha \lambda t_{z}k_{z}\left( k_{x}\sin \phi +k_{y}\cos \phi
\right) .
\end{equation}%
Solving $f_{x}=f_{z}=f_{y}=0$, we obtain the Fermi surface.

When $\phi =\pi /4$ or $-\pi /4$, there emerge a loop node along the $%
k_{x}=k_{y}$ plane or the $k_{x}=-k_{y}$ plane, and two zero-energy points
emerge at the two points $\left( k_{x},k_{y},k_{z}\right) =\left( \pm
k_{c},\pm k_{c},0\right) $ or $\left( \pm k_{c},\mp k_{c},0\right) $ with $%
k_{c}=\sqrt{m/2t}$. By expanding the Hamiltonian around these points the
dispersion relation is found to be linear. For instance, at $\phi =\pi /4$
it reads 
\begin{equation}
H=\pm k_{c}\left( tk_{y}^{\prime \prime }\sigma _{x}-2\alpha k_{z}\sigma
_{y}-2\lambda k_{x}^{\prime \prime }\sigma _{z}\right)
\end{equation}%
with $k_{x}^{\prime \prime }=k_{x}^{\prime }+k_{y}^{\prime }$, $%
k_{y}^{\prime \prime }=k_{x}^{\prime }-k_{y}^{\prime }$, $k_{x}^{\prime
}=k_{x}\mp k_{c}$, $k_{y}^{\prime }=k_{x}\pm k_{c}$. Hence, they are Weyl
point nodes carrying the unit monopole charge: See Figs.\ref{FigBerry}(a4)
and (b4). Unless $\phi =\pm \pi /4$, only Weyl point nodes appear as in Fig.%
\ref{FigBerry}(a5) and (b5).

The lattice Hamiltonian of the trigonal symmetric model with $N=3$ is given
by%
\begin{align}
f_{x}& =t\left( \cos k_{x}+\sum_{\eta =\pm 1}\cos \frac{-k_{x}+\eta \sqrt{3}%
k_{y}}{2}\right) +t_{z}\cos k_{z}-m,  \notag \\
f_{z}& =\frac{\lambda }{2}\sin k_{x}\left( \cos k_{x}-\cos \sqrt{3}%
k_{y}\right) .
\end{align}%
The photoinduced term reads in the continuum theory as 
\begin{equation}
f_{y}=\alpha \frac{3\lambda t_{z}}{4}k_{z}\left( \left(
k_{x}^{2}-k_{y}^{2}\right) \sin \phi +2k_{x}k_{y}\cos \phi \right) .
\end{equation}%
When $\phi =\pi /2$, $\pi /2\pm 2\pi /3$, a loop node and four point nodes
emerge as in Figs.\ref{FigBerry}(c4) and (d4), and otherwise only points
modes emerge. Namely, a loop node emerge only when the direction of
photoirradiation is perpendicular to the loop node. Figs.\ref{FigBerry}(c5)
and (d5). 
\begin{figure}[t]
\centerline{\includegraphics[width=0.50\textwidth]{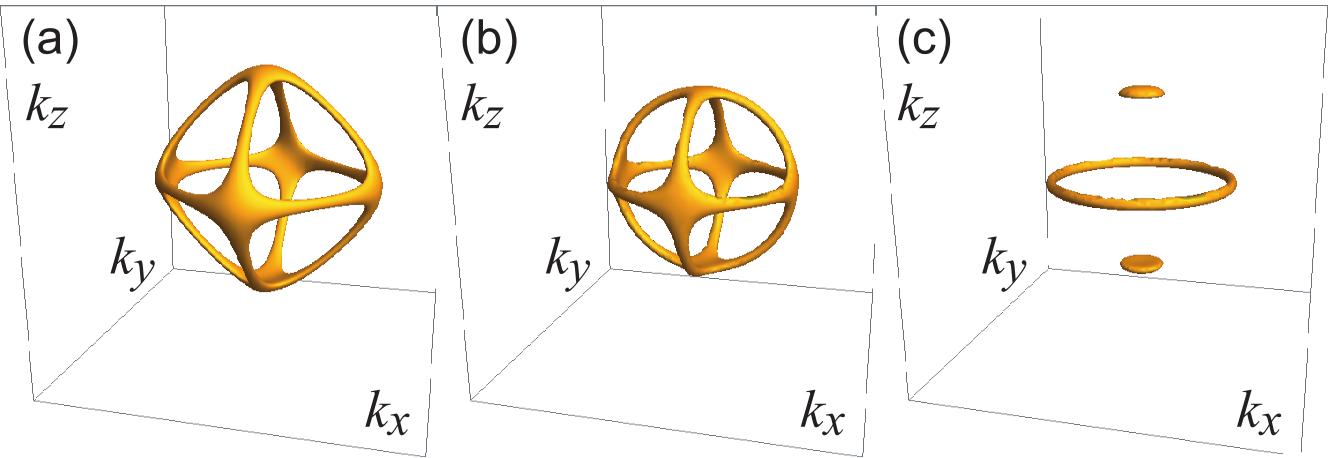}}
\caption{Bird's eye's view of the almost zero-energy surfaces of the
Hamiltonian with the cubic symmetry for (a) the lattice model, (b) the
continuum model, (c) the continuum model with photoirradiation along the $%
k_{z}$ axis. Each loop node carries the unit Berry magnetic flux in
(a)--(c). The point nodes at the north and south poles carry the $\pm 2$
units of Berry monopole charges. }
\label{FigCubic}
\end{figure}

\textit{Cubic symmetric model: }Finally, we present a simple realization of
a lattice model with the cubic symmetry by taking 
\begin{align}
f_{x}& =t\left( \cos k_{x}+\cos k_{y}+\cos k_{z}\right) -m^{\prime },  \notag
\\
f_{z}& =\lambda ^{\prime }\sin k_{x}\sin k_{y}\sin k_{z},  \label{EqB}
\end{align}%
where we have set $g(k_{z})=\sin k_{z}$ in (\ref{EqA}). We illustrate the
Fermi surface of the lattice model and the continuum model in Fig.\ref%
{FigCubic}(a) and (b). Photoirradiation applied along the $z$ direction
induces the term 
\begin{equation}
f_{y}=\alpha \lambda tk_{z}\left( k_{x}^{2}-k_{y}^{2}\right)
\end{equation}
in the continuum theory. Solving $f_{x}=f_{z}=f_{y}=0$, we obtain a loop
mode given by $k_{x}^{2}+k_{y}^{2}=2(3t-m)/t$ and $k_{z}=0$. Additionally
anisotropic double-Weyl points emerge at the north and south poles, carrying
the monopole charge $\pm 2$. We illustrate the Fermi surface in Fig.\ref%
{FigCubic}(c).

\textit{Discussion:} Crossing-line nodal semimetals with the cubic symmetry
are realizable in CaTe and Cu$_3$PdN according to recent first principles
calculations in Ref.\cite{CaTe} and Ref.\cite{Yu,Kim}, respectively. It is
also shown that LaN has a crossing line node, which is topologically
identical to the cubic symmetric model in Ref.\cite{Zeng}. Furthermore, a
hexagonal hydride, YH$_{3}$, has a crossing-line nodes with $N=3$, as shown
in Ref.\cite{Yamakage}. It is an interesting problem to search further
materialization of crossing-line node semimetals.

The author is very much grateful to N. Nagaosa for many helpful discussions
on the subject. This work is supported by the Grants-in-Aid for Scientific
Research from MEXT KAKENHI (Grant Nos.JP17K05490 and 15H05854). This work
was also supported by CREST, JST (Grant No. JPMJCR16F1).


\end{document}